\documentclass[12pt]{article}
\usepackage{graphicx}
\usepackage[bookmarksnumbered,colorlinks,plainpages]{hyperref}

\usepackage{url}
\usepackage{amsmath}
\usepackage{amscd}
\usepackage{subfig}
\begin{document}

\title{The effect of interdependence on the percolation of interdependent networks}

\author{J. Jiang$^{1}$, W. Li$^{2}$ and X. Cai$^{2}$ \\
{\it $^{1}$Research Center of Nonlinear Science and}\\
{\it College of Mathematics and Computer Science of Wuhan}\\
{\it Textile University, Wuhan, 430200, P R, China }\\
{\it $^{2}$Complexity Science Center, Institute of Particle}\\
{\it Physics, Hua-Zhong (Central China) Normal University, Wuhan}\\
{\it 430079, P R, China }}

\date{}

\maketitle

\begin{abstract}
Two stochastic models are proposed to generate a system composed
of two interdependent scale-free (SF) or Erd\H{o}s-R\'{e}nyi (ER)
networks where interdependent nodes are connected with exponential
or power-law relation, as well as different dependence strength,
respectively. Each subnetwork grows through the addition of new
nodes with constant accelerating random attachment in the first
model but with preferential attachment in the second model. Two
subnetworks interact with multi-support and undirectional
dependence links. The effect of dependence relations and strength
between subnetworks are analyzed in the percolation behavior of
fully interdependent networks against random failure, both
theoretically and numerically, and as a result, for both
relations: interdependent SF networks show a second-order
percolation phase transition and increased dependence strength
decreases the robustness of the system, whereas, interdependent ER
networks show the opposite results. In addition, power-law
relation between networks yields greater robustness than
exponential one at given dependence strength.

\end{abstract}


\section{Introduction}
Nowadays, with enhanced development of modern technology, the
interaction between networks becomes increasingly intensive and
complicated \cite{lkk, v,kp}. Examples of interdependent networks
are ubiquitous and include, subway network and airport network in
transportation system, bank network and company network in economy
system, communication network and power grid network in
infrastructure system, and so forth. In these interdependent
networks, the failures of nodes in one subnetwork generally will
lead to the failure of dependent nodes in the other subnetworks
\cite{bppsh, gbsh,sbhs,hgbhs,pbh2,bph}. This may happen
recursively and might lead to a cascade of failures. Understanding
how robustness is affected by the interdependence between
subnetworks becomes one challenge when designing resilient
systems. Very recently, several studies presented a theoretical
framework for studying the process of cascading failures in
interdependent networks and showed that interdependencies
significantly increase the vulnerability of the entire networks to
random attack \cite{pbh,hkch,gbhs2,gbhs3}. In addition, the
first-order phase transition presented in interdependent networks
is totally different from the second-order phase transition
occurred in isolated network.

Most existing studies have focused almost exclusively on random
interdependent networks in which the interdependent nodes are
randomly connected, which is at odds with real complex systems.
Taking the Italian power grid and communication networks as an
example \cite{bppsh,pbh,ritmps}, it is very common that a central
communication station depends on a central power station and vice
versa. Similarly, well-connected seaports are found more likely to
depend on well-connected airports in Ref.\cite{pridh} where
positive correlation exists between the interaction of
subnetworks. Based on this feature, interdependence with
correlation, not random, has attracted much attention in the
robustness of interdependent networks currently. Parshani
\cite{pridh} and Cho \cite{cck} have shown similar result that the
positive correlated interdependence enhances the robustness of
networks, respectively. Buldyrev et.al \cite{bsc} have
analytically investigated the situation with one simple
correlation that all pairs of interdependent nodes have the same
degree. In addition, Ref.\cite{sbhs} and Ref.\cite{phds} have
discussed the interdependence relation represented by Poisson
distribution and power-law distribution in stochastic models,
respectively. Furthermore, the effect of the dependence strength
between subnetworks also plays the key role in the percolation of
interdependent networks. Ref.\cite{pbh} has found that when the
dependence strength is reduced, the percolation transition becomes
second-order transition at a critical coupling strength, which
enhanced the robustness of the system. How and to what extent the
relation of interdependence between subnetworks might influence
the entire system's structure and function are still not well
known.

In present work, discussing the effect of different dependence
relation and dependence strength on the robustness of interacting
system under random attack is our focus and motivation. Two types
of relations are generated by two stochastic growing network
models whereby the origin of relations is explained. One is that
interdependent nodes randomly depend with each other with
exponential degree distribution, the other is that they
preferentially depend with each other with power-law degree
distribution. In addition, two interdependent scale-free (SF) and
Erd\H{o}s-R\'{e}nyi (ER) networks are also created in these two
models, respectively. Besides, the influences of dependence
relations and coupling strength of multi-support, undirectional
dependence links on the robustness of networks are theoretically
analyzed and simulated. As a result, it is found that, (1) two
different interdependence links could be generated by the addition
of dependence links; (2) for interdependent SF networks and ER
networks, different types of phase transition and opposite effects
of dependence strength are presented; (3) for the effect of
interdependence, power law distribution of dependence degree
yields higher robustness than exponential one with given
dependence strength.

\section{The first model}
In both two models, there are two types of links among the nodes:
connectivity links (intra-links in each subnetwork) that enable
the nodes to function cooperatively as a network, and dependence
links (cross-links between subnetworks) that bind the failure of
one subnetwork node to the failure of other subnetwork nodes.
These two kinds of links correspond to two kinds of degree of each
node in networks, connectivity degree ($k_{con}$) and dependence
degree ($k_{dep}$), respectively. The first model of two
interdependent scale free (SF) networks is built by the following
considerations.

Initially, both subnetworks A and B contain $m_0$ nodes and $n_0$
connectivity links, without dependence links between subnetworks.
At each time step $t$, two new nodes are introduced
simultaneously, one belonging to subnetwork A and the other
belonging to subnetwork B. The new node joining to subnetwork A
with $m_A$ links added, preferentially attaches $1-q_A$ fraction
of its links as connectivity links to pre-existing nodes in
subnetwork A. The rate of acquiring a link replies on the degrees
of pre-existing nodes in subnetwork A. And then this new node
randomly or preferentially attaches $q_A$ fraction of links as
dependence links to pre-existing nodes in subnetwork B. In other
words, the connectivity degree and the dependence degree of the
new node joining to subnetwork A are equal to $m_A(1-q_A)$ and
$m_Aq_A$ at time step $t$ through different addition methods,
respectively. The similar process is executed when a new node
joins to subnetwork B, where the new node has $m_B$ links added
from which $1-q_B$ fraction of them randomly connect pre-existing
nodes in subnetwork B and $q_B$ fraction of them randomly or
preferentially connect to pre-existing nodes in subnetwork A, and
its connectivity degree and dependence degree are equal to
$m_B(1-q_B)$, $m_Bq_B$, respectively. $q_A$ and $q_B$ are defined
as the strength of dependence between two subnetworks. Larger
$q_A(q_B)$ means the more dependence links between subnetworks or
the more intensively two subnetworks depend on each other. The
process ends when the size of both subnetworks increases up to
$N$. In fact, through this model, the subnetworks A and B
generated are equivalent to the classical random graph studied by
Barab\'{a}si-Albert with power-law degree distribution
($p(k_{con})$), and thereby named two interdependent SF networks.
Two dependence relations between interdependent nodes are
represented by the degree distribution of dependence links
$p(k_{dep})$. One is exponential distribution with general form
\cite{soj}
\begin{equation}\label{expdistribution}
p(k_{dep})=\frac{1}{mq+1}(\frac{mq}{mq+1})^{k_{dep}-mq},
\end{equation}
$mq\geq1$, with random dependence between subnetworks with
supposition $m_A=m_B=m$ and $q_A=q_B=q$, and the other is
power-law distribution $p(k_{dep})\sim k^{-3}_{dep}$ \cite{ba}
with preferential dependence between subnetworks.

\begin{figure} [!h]
\includegraphics[scale=.2]{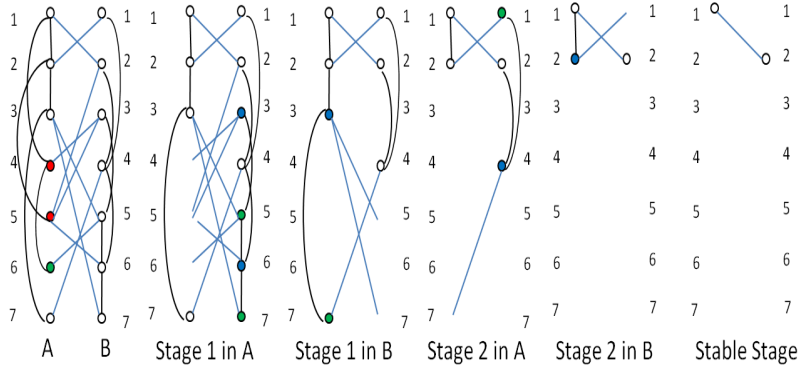}
 \centering\caption{(Color online) Description of the process of
cascading failures in two fully interdependent networks. Black
lines represent connectivity links, and blue lines represent
dependence links. The white nodes represent the survival nodes,
the red nodes represent the attacked nodes, the green nodes
represent the ones separate from the giant component of networks,
the blue nodes represent the ones without dependence links.
Initially, nodes 4 and 5 (red in \textbf{A}) are attacked and
removed from subnetwork A. \textbf{Stage 1 in A}: node 6 (green in
\textbf{A}) is removed because it does not belong to the giant
component of subnetwork A. \textbf{Stage 1 in B}: nodes 3 and 6
(blue in \textbf{Stage 1 in A}) are removed because they lose all
their dependence links, nodes 5 and 7 (green in \textbf{Stage 1 in
A}) are removed because of separation from the giant component of
subnetwork B. The similar process is carried out in stage 2. Note
that: the failure of node 4 (blue in \textbf{Stage 2 in A})
results in two giant components with same size in subnetwork B. In
this case, we randomly choose one giant component to fail as node
1 marked in green in \textbf{Stage 2 in A}. After two stages, the
interdependent networks reaches a stable state, since no further
failure occurs in networks.} \label{pic}
\end{figure}

The iterative process of cascading failures is initiated by
randomly removing a fraction $1-p$ of nodes from subnetwork A and
all edges linked to them. When nodes in subnetwork A fail, the
interdependent nodes in subnetwork B also fail. Specially, we
suppose that only the nodes in the giant component with at least
one dependence link remain functional, which leads to the further
failure in the first subnetwork. This dynamic process ends with no
further node failure in the system. The cascade of failures in
small interdependent networks with $N=7$ is demonstrated in
Fig.\ref{pic}.

The dynamics of cascading failures is performed as following and
$g_A$ and $g_B$ are defined as the fraction of nodes belonging to
the giant component of subnetwork A and B, respectively
\cite{gbsh}. After the initial removal of $1-p$ fraction of nodes
in subnetwork A, the remaining fraction of subnetwork A nodes is
$\psi_1'=p$. The remaining functional part of subnetwork A
contains a fraction $\psi_1=\psi_1' g_A(\psi_1')$ of the network
nodes. Since the number of dependence links $k_{dep}^B$ of each
node in subnetwork B is multiple and a random number, the
probability that a node in subnetwork B has no dependence links in
subnetwork A is $\mu_1^B =
\sum_{k^B_{dep}}p^B(k^B_{dep})(1-\psi_1)^{k^B_{dep}} =
\tilde{G}^B(1-\psi_1)$ ($\tilde{G}^B$ the generating function of
degree distribution $p^B(k_{dep}^B)$). Accordingly, the remaining
fraction of subnetwork B is $\phi_1'=1-\mu_1^B$, and the fraction
of nodes in the giant component of subnetwork B is $\phi_1=\phi_1'
g_B(\phi_1')$. Following this method, the sequence of giant
components, $\psi_n$ and $\phi_n$,  and that of the remaining
fractions of nodes, $\psi_n'$ and $\phi_n'$, at each stage of the
cascading failures are constructed as following:
\begin{eqnarray}\label{finalequs}
\psi_1'&=&p, \qquad \psi=\psi_1'g_A(\psi_1'),\qquad \phi_0'=1,\nonumber\\
\phi_1'&=&1-\tilde{G}^B(1-p g_A(\psi_1')), \qquad \phi_1=\phi_1'
g_B(\phi_1'),\nonumber\\  &\ldots,&\\
\psi_n'&=&p[1-\tilde{G}^A(1-g_B(\phi_{n-1}'))],\ \psi_n=\psi_n'g_A(\psi_n'),\nonumber\\
\phi_n'&=&1-\tilde{G}^B(1-pg_A(\psi_n')),\qquad
\phi_n=\phi_n'g_B(\phi_n').\nonumber
\end{eqnarray}
The final size of each subnetwork at the end of the cascade
process can be represented by $\psi'_n$, $\phi'_n$ at the limit of
$n\rightarrow \infty$. This limit satisfies the equations $\psi'_n
= \psi'_{n+1}$ and $\phi_n'=\phi_{n+1}'$ since the cluster is not
further fragmented. An exact analytical solution can be obtained
using the formalism of generating functions. According to
Refs.\cite{n,sbckhs}, the generating functions of the degree
distributions of subnetworks A and B,
$G_{A0}(x)=\sum_{k^A_{con}}p^A(k^A_{con})x^{k^A_{con}}$ and
$G_{B0}(x)=\sum_{k^B_{con}}p^B(k^B_{con})x^{k^B_{con}}$ are
introduced. Analogously, the generating functions of the
underlying branching processes, $G_{A1}(x)=G_{A0}'(x)/G_{A0}'(1)$
and $G_{B1}(x)=G_{B0}'(x)/G_{B0}'(1)$ are also introduced. As the
random removal of fraction $1-p$ of nodes will change the degree
distribution of the remaining nodes, so the generating functions
of the new distribution are equal to generating functions of the
original distribution with the argument $x$ replaced by $1-p(1-x)$
\cite{sbbhs}. The fraction of nodes that belong to the giant
component after the removal of $1-p$ nodes is
$g_A(p)=1-G_{A0}[1-p(1-f^A)]$, where $f^A=f^A(p)$ satisfies a
transcendental equation $f^A=G_{A1}[1-p(1-f^A)]$.

As the theoretical analysis of generating function with power-law
distribution is not available in the first model, we just present
the numerical result here with $N=10^4$, $m=5$ in simulations.

\begin{figure}[!h]
\includegraphics[scale=.45]{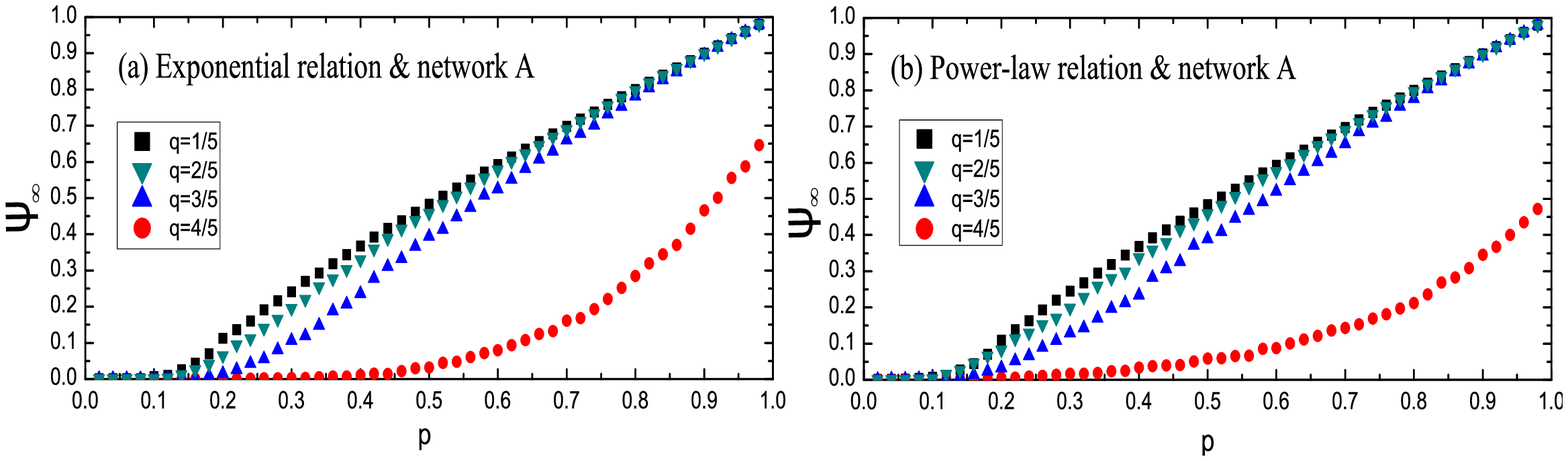}
\centering \caption{(Color online) The dependence of giant
components $\psi_{\infty}$ of subnetwork A with different
dependence strength and relations on $p$ at $N=10^4$ and $m=5$.
For two relations, $\psi_{\infty}$ changes continuously from a
finite value to zero at critical threshold $p_c$, characterizing
the second-order phase transition occurred in the system, and the
value of $p_c$ increases with the increasing of $q$ in both
cases.} \label{Pre_Pre_psi_p}
\end{figure}

Fig.\ref{Pre_Pre_psi_p} shows the effect of different dependence
relations, exponential and power-law relation in the function of
$\psi_{\infty}$, the fraction of nodes in giant component of
subnetwork A, after a random attack with different dependence
strength $q$. We find two common points for both relations: (1)
$\psi_{\infty}$ has similar tendency against $p$ with different
dependence strength $q$. It smoothly decreases to zero at critical
point $p_c>0$ characterizing a second-order phase transition. This
result differs from the general known result the first-order phase
transition discovered in coupled networks \cite{bppsh,pbh}; (2)
with the increasing of dependence strength $q$, the value of
critical point $p_c$ increases, which implies the decreasing of
resilience of networks. The potential reasons for this may be that
since the sum of connectivity link and dependence link per new
node at each time step is constant, the larger dependence strength
$q$ means the less the connectivity links (the smaller mean
connectivity degree) and the more the dependence links of each
node in each subnetwork, or stronger interdependence between
subnetworks. Smaller mean connection degree quickens the
fragmentation of individual network and hubs in one network can
depend on weak (low-degree) nodes in the other network and
vice-versa, and then the strong interdependence leads to
accelerated cascades of failures.

\begin{figure}[!h]
\includegraphics[scale=.4]{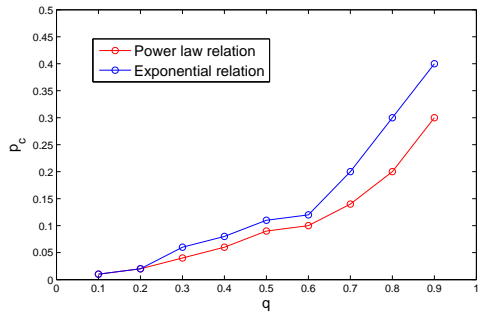}
\centering \caption{(Color online) The tendency of critical
threshold $p_c$ on dependence strength $q$ with two dependence
relations. When $q$ is close to zero, interdependent SF networks
becomes single scale-free network where the critical threshold
$p_c=0$ is obtained under random attack in percolation. For small
$q$ around 0.2, networks with both relations have the same $p_c$
under cascading failures. For $q>0.2$, networks with power-law
relation between dependent nodes have smaller $p_c$ than those
with exponential one, which indicates that power-law distribution
of dependent degrees yields grater robustness of system than
exponential one. } \label{pre_q_pc}
\end{figure}

The discrepancy of effects caused by two relations is shown in
Fig.\ref{pre_q_pc}. The critical point $p_c$ is an increasing
function of dependence strength $q$. When $q$ is close to zero
corresponding to the extreme case that there is no interdependence
between subnetworks, $p_c$ attends to zero and goes back to the
classical case that the single scale-free network has critical
percolation value $p_c=0$ under random failure. For weak
dependence strength $q$ around $q=0.2$, the same $p_c$ is found
for two dependence relations. In the range of $q>0.2$, the value
of $p_c$ for power-law relation is always smaller than that for
exponential relation, which demonstrates that power-law
distribution of dependence degree yields greater robustness than
exponential one when certain dependence strength $q$ is given.
This result could be attributed to the possibility that power-law
relation between the dependent nodes could suppress the phenomenon
of hubs in one network becoming vulnerable by being dependent on
weak nodes in the other network, when the dependence strength
arrives at certain critical threshold. In addition, this finding
strengthens the conclusions of recent studies \cite{pridh,cck,bsc}
that coupled networks with positively correlated degrees of
dependent nodes are always more robust than randomly coupled
networks.

\section{The second model}
Two interdependent ER networks is generated in second model, which
is the difference from the first model. The common place between
two models is that subnetworks depend on each other with two
identical relations. In second model, more attention is paid on
the theoretical analysis of the effect brought by interdependence.
This model is constructed as following.

Initially, both subnetworks A and B contain $m_0$ nodes and $n_0$
connectivity links, without dependence links between subnetworks.
At each time step $t$, two new nodes, one belonging to subnetwork
A and the other belonging to B, are introduced simultaneously.
Connectivity links will be created by the scenario of constant
acceleration (See Ref.\cite{soj} for details). It is processed as
follows: for subnetwork A, connectivity links between the new node
and pre-existing nodes are established randomly with probability
$s$ satisfying the requirement that the expected number of links
for the new node is equal to $st$. For the addition of dependence
links, there are two approaches like those in the first model:
randomly or preferentially connect the new node belonging to
subnetwork A to $DL$ pre-existing nodes in subnetwork B, which
will generate the exponential and power-law distribution of
dependence degrees, respectively. $DL$ is defined as the strength
of dependence between two subnetworks like $q$ in the first model.
For subnetwork B, similar process is carried out in the creation
of connecting and dependence links. Through this model, the
subnetworks A and B generated are equivalent to the classical
random graph studied by Erd\H{o}s-R\'{e}nyi with Poisson degree
distribution, and thereby named two interdependent ER networks.
When the size of subnetworks A and B increases to $N$, the process
of building two interdependent ER networks is concluded.

According to the dynamics of cascading failures described in the
first model, for interdependent ER networks, the problem can be
solved explicitly, since $G_0(x)$ and $G_1(x)$ have the same
simple form $G_0(x)=G_1(x)=e^{\langle k\rangle(x-1)}$ \cite{n}.
Supposing that the average degree of subnetwork A is $\langle
k\rangle=a$, and for subnetwork B, one gets $\langle k\rangle=b$.
Thus, from $g_A(\psi_{\infty}')=1-f^A$ and
$g_B(\phi_{\infty}')=1-f^B$, both $
g^A(\psi_{\infty}')=1-e^{-a\psi_{\infty}}$ and $
g^B(\phi_{\infty}')=1-e^{-b\phi_{\infty}}$ are reduced. According
to the definitions in Eqs.(\ref{finalequs}) at the limit of
$n\rightarrow \infty$, the giant components of subnetwork A and B
with generating functions of dependence relations $\tilde{G}^A$
and $\tilde{G}^B$ at the stable state are obtained:
\begin{eqnarray}
\psi_{\infty}&=& p[1-\tilde{G}^A(e^{-b\phi_{\infty}})](1-e^{-a\psi_{\infty}}),\label{final1}\\
\phi_{\infty}&=&
[1-\tilde{G}^B(1-p(1-e^{-a\psi_{\infty}}))](1-e^{-b\phi_{\infty}}).\label{final2}
\end{eqnarray}
(1) In the case of exponential dependence relation between
subnetworks, according to the definition of $\tilde{G}^A$,
$\tilde{G}^B$ and Eq.(\ref{expdistribution}), Eqs.(\ref{final1})
and (\ref{final2}) become:
\begin{eqnarray}
\psi_{\infty} & =& p[1-\frac{1}{-DL+e^{b\phi_{\infty}}+DL
e^{b\phi_{\infty}}}] (1-e^{-a\psi_{\infty}}),\label{psi}\\
\phi_{\infty} &=&
[1+\frac{-e^{a\psi_{\infty}}-p+pe^{a\psi_{\infty}}}{e^{a\psi_{\infty}}-DLp+DLpe^{a\psi_{\infty}}}]
(1-e^{-b\phi_{\infty}}).\label{phi}
\end{eqnarray}
(2) In the case of power-law dependence relation between
subnetworks, based on the simulation result, the degree
distributions of dependence degrees with different link addition
$DL$ are found to have the same factor of $\sim k_{dep}^{-3}$ over
the central range of degree and have various minimum degree
$k_{dep}^{min}$. Hence the formation of degree distribution
$p(k_{dep})=(k_{dep}^{min}/k_{dep})^2-[k_{dep}^{min}/(k_{dep}+1)]^2$,
behaving asymptotically as $2(k_{dep}^{min})^2/k_{dep}^3$, is used
in the calculation of $\tilde{G}^A$ and $\tilde{G}^B$. In
addition, $DL=1,2,8$ correspond to $k_{dep}^{min}=1,2,8$ in the
degree distributions, respectively. Similarly, along the
definition of $\tilde{G}^A$ and $\tilde{G}^B$, Eqs.(\ref{final1})
and (\ref{final2}) become:
\begin{equation}
\begin{aligned}
&\psi_{\infty} = p [\text{PolyGamma}[2,k_{dep}^{min}]+2(e^{-b
\phi_{\infty} })^{k_{dep}^{min}}\\  &\text{HurwitzLerchPhi}
[e^{-b\phi_{\infty} },3, k_{dep}^{min}]]
(1-e^{-a \psi_{\infty}})\\ &/{\text{PolyGamma}[2,k_{dep}^{min}]},\\
&\phi_{\infty} = [\text{PolyGamma}[2,k_{dep}^{min}]+2
[1+(-1+e^{-a\psi_{\infty} }) p]^{k_{dep}^{min}}
\\& \text{HurwitzLerchPhi}
 [1+(-1+e^{-a \psi_{\infty} })
p,3,k_{dep}^{min}] ] (1-e^{-b \phi_{\infty} })\\
&/{\text{PolyGamma}[2,k_{dep}^{min}]},
\end{aligned}
\end{equation}
where $\text{PolyGamma}[n,z]$ gives the $n^{th}$ derivative of the
digamma function $x^{n}(z)$, $x(z)=\Gamma'(z)/\Gamma(z)$, and
$\text{HurwitzLerchPhi}[z,s,a]$ gives the Hurwitz-Lerch
transcendent $\Phi(z,s,a)$
($\Phi(z,s,a)=\sum_{k=0}^{\infty}z^k(k+a)^{-s}$).

In the first case (1), in the limit of $DL\rightarrow \infty$, the
giant component of two interdependent ER networks will not depend
on each other and the percolation theory of single network
$\psi_{\infty}=p(1-e^{-a \psi_{\infty}})$ is recovered, which is
comparable with the result of reference \cite{sbhs} where the
Poisson degree distribution was given between interdependent
nodes. The solutions of system with Eqs.(\ref{psi}) and
(\ref{phi}) can be graphically presented by the intersection of
the curves $\phi_{\infty}(\psi_{\infty})$ and
$\psi_{\infty}(\phi_{\infty})$ . The trivial solutions correspond
to $\psi_{\infty}=\phi_{\infty}=0$ and the nontrivial solutions in
the critical case can be found from the tangential condition
$\frac{\text{d$\psi $}_{\infty }\left(\phi _{\infty
}\right)}{\text{d$\phi $}_{\infty }}\frac{\text{d$\phi $}_{\infty
}\left(\psi _{\infty }\right)}{\text{d$\psi $}_{\infty }}=1$,
corresponding to the single point of two curves. Together with
Eqs.(\ref{psi}) and (\ref{phi}), the critical value of the
parameters $a,b,DL,p$ can be reduced when three of them are fixed.
Here, with the assumption of $a=b$, we get the expression of
critical threshold $p_c$ above which two interdependent ER
networks have non-zero giant components:
\begin{equation}\label{pc}
p_c=\frac{5a+7a DL-4DL+2a DL^2-2DL^2}{2a(1+DL)(a+(a-1)DL)}.
\end{equation}
When $a$ is fixed and $DL\rightarrow \infty$, the above equation
will become: $p_c=1/a$, which is the critical threshold of random
percolation for single ER network \cite{gbsh}. In the second case
(2), the exact theoretical results for networks with power-law
dependence relation are not available, so the numerical simulation
results will be given below.

\subsection{Numerical simulations}
\begin{figure}[h!]
\includegraphics[scale=.05]{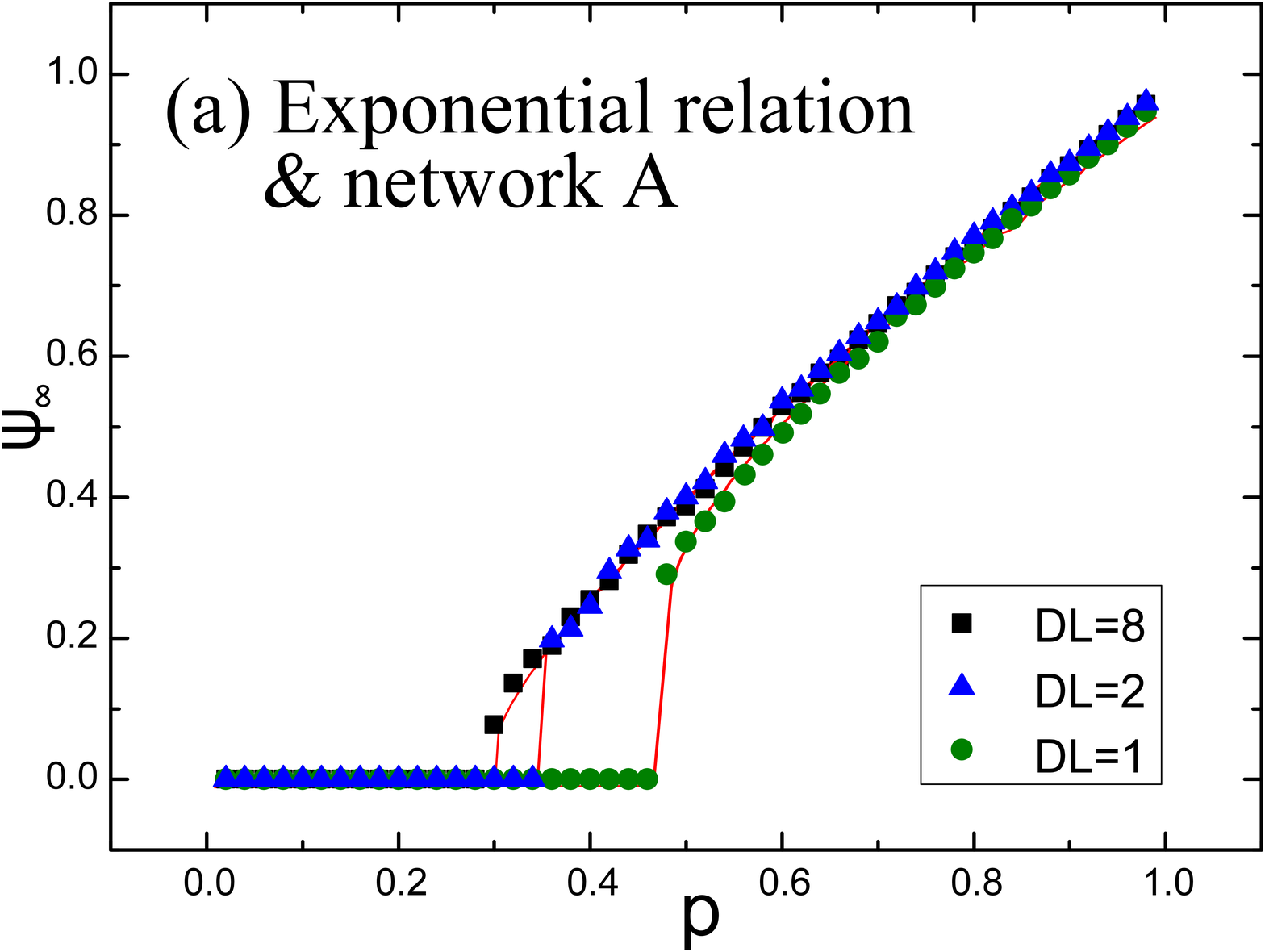}
\includegraphics[scale=.05]{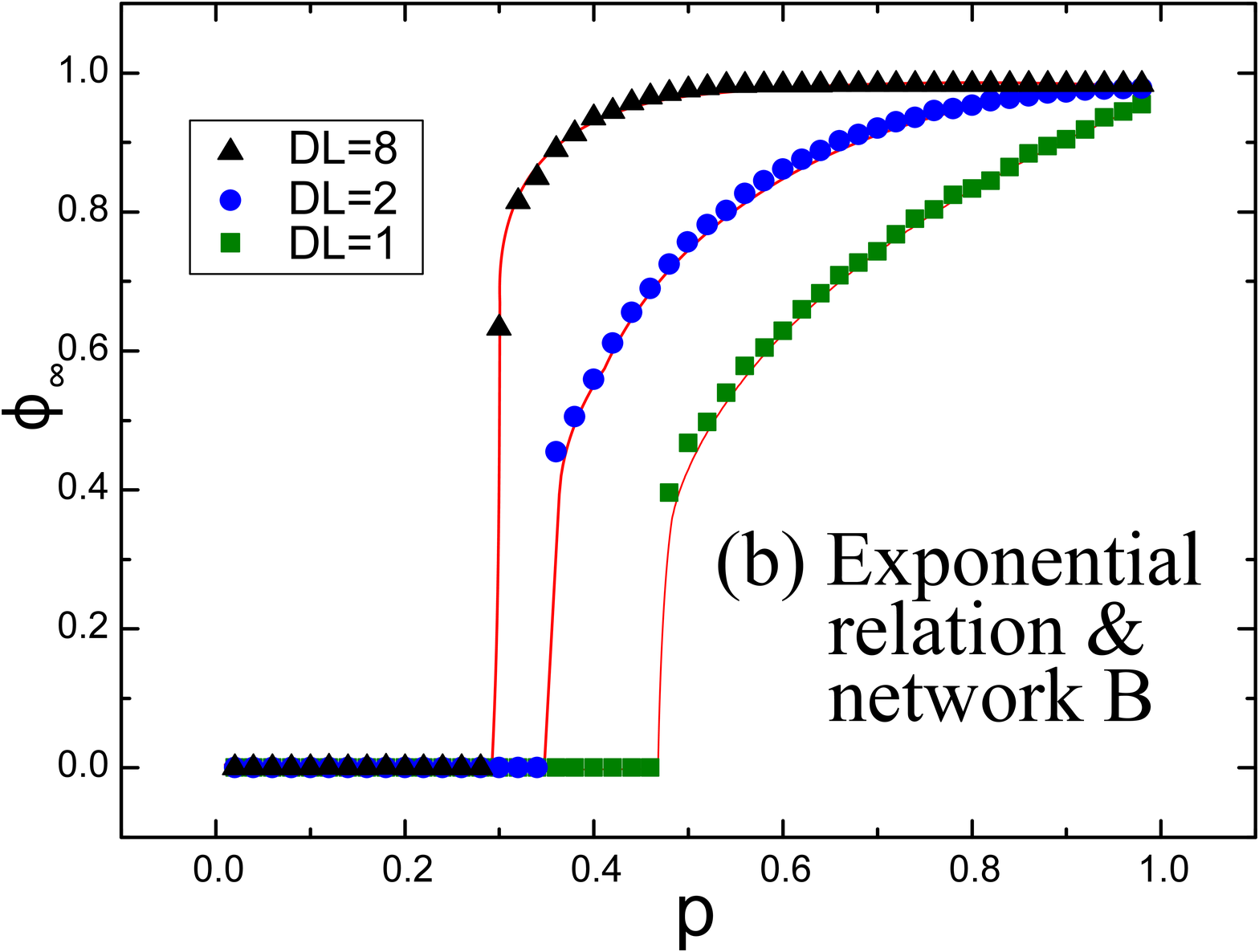}
\includegraphics[scale=.05]{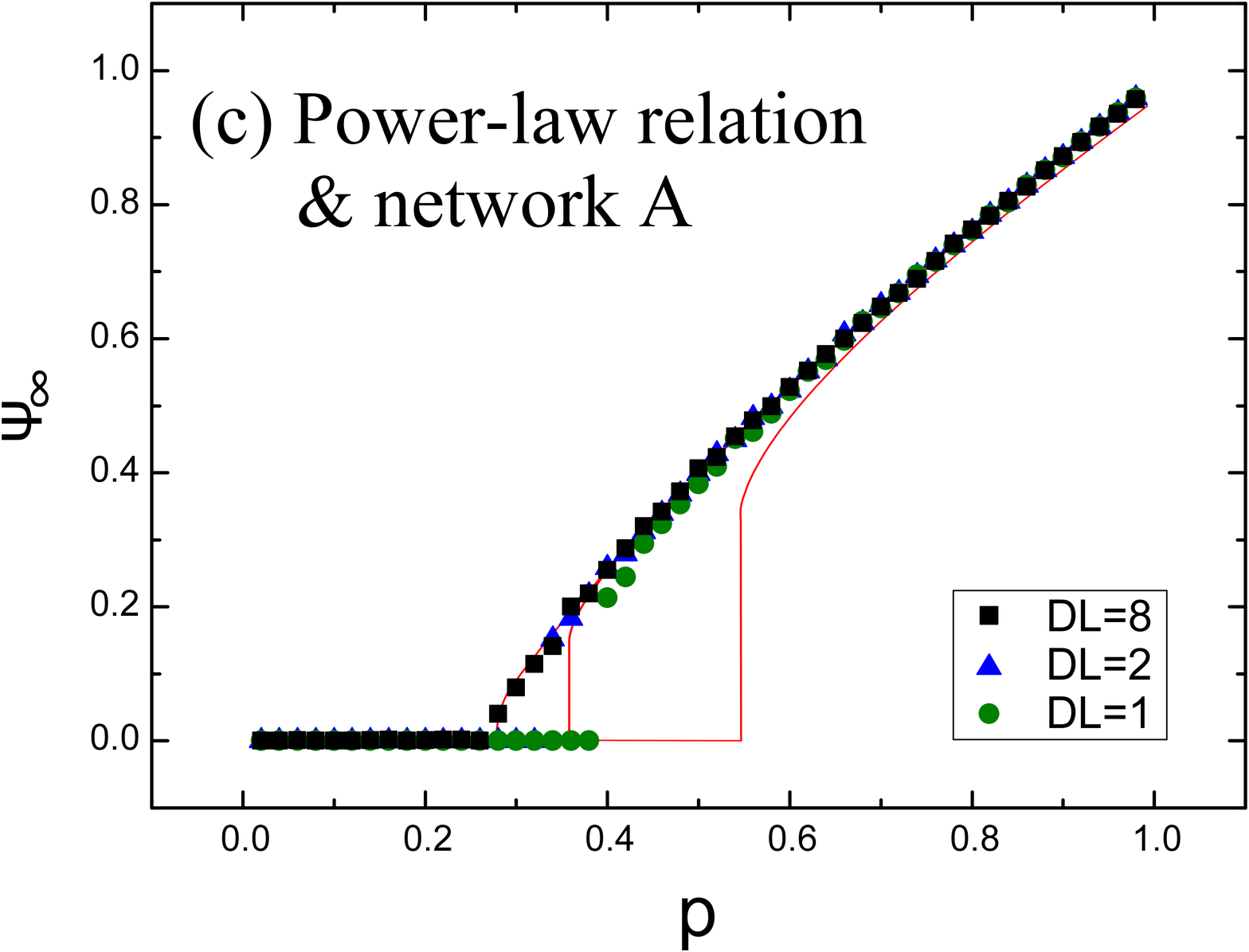}
\includegraphics[scale=.05]{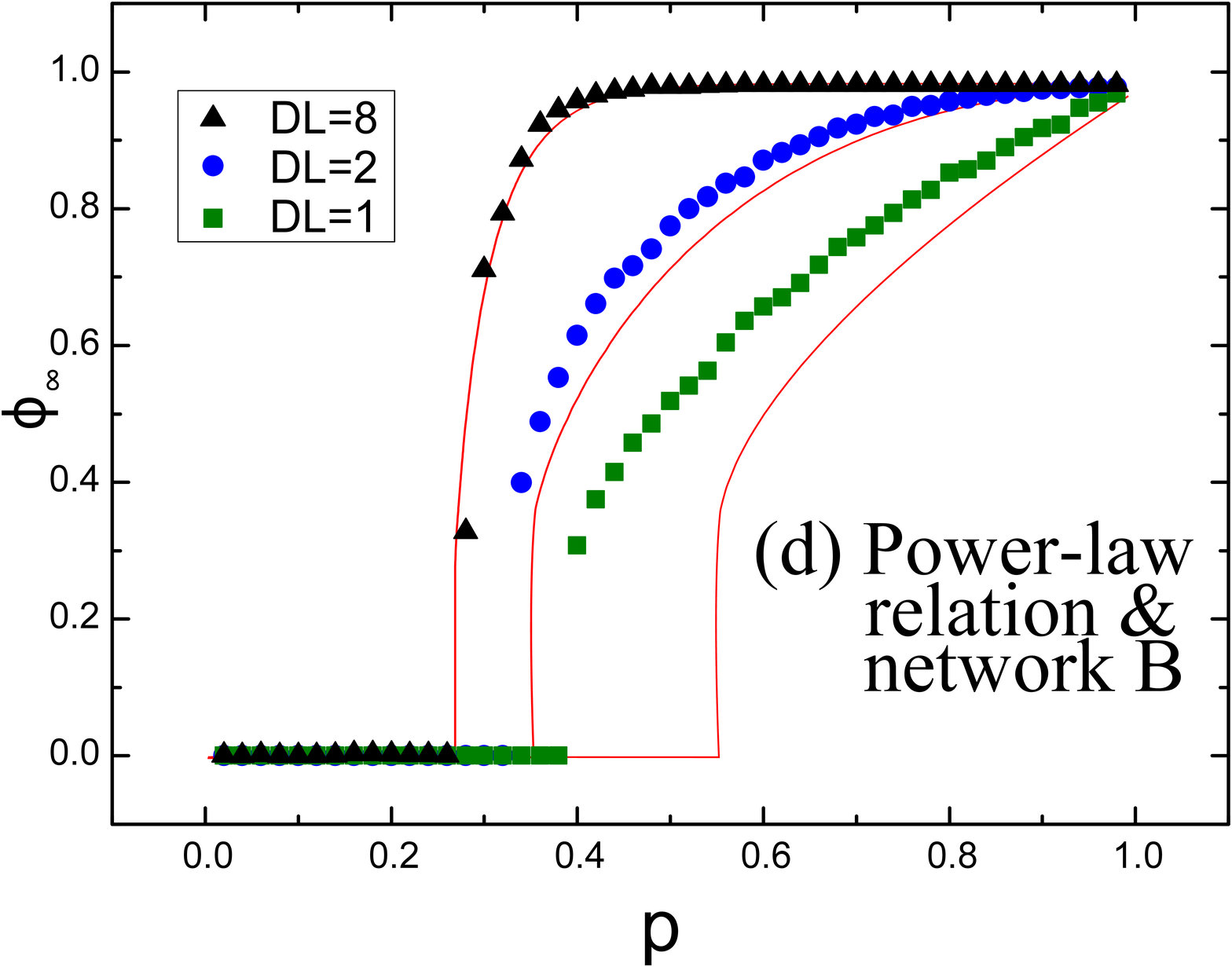}
\centering \caption{(Color online) The dependence of giant
components $\psi_{\infty}$ and $\phi_{\infty}$ of two
interdependent ER networks with two different dependence relations
on $p$. In all cases, $N=10^5$, $a=b=4$, $DL=1,2,8$. In panels (a)
and (b), for the system with exponential dependence relation, as
$DL$ increases, the giant components discontinuously change from
finite value to zero at critical threshold $p_c\rightarrow 0.25$
that is the critical point of random percolation of a single ER
network for infinite $DL$. The simulation results (symbols) agree
well with analytical results (red lines). In panels (c) and (d),
for the system with power-law dependence relation, similar
tendency is found in the dependence of giant components of both
networks on parameter $p$. There is a little deviation between the
analytical results and simulations in the case of $DL=1$. In
addition, as $DL\rightarrow \infty$, $\phi_{\infty}$ in panels (b)
and (d) converges to a Heaviside step function. } \label{ER_Psi_p}
\end{figure}

In this section, the theoretical results discussed in above
section are compared with results of numerical simulations. In all
simulations, we have $N=10^5$ and $a=b=4$. In Fig.\ref{ER_Psi_p},
the giant components of two interdependent ER networks with two
dependence relations as a function of $p$, the fraction of nodes
in subnetwork A needed to be preserved at the beginning of the
cascading failures is shown. In panels (a) and (b), for
exponential dependence relation, as $DL$ increases, the critical
value of $p^A_c$ ($p^B_c$) is close to 0.25 eventually, the
critical threshold value of random percolation of a single ER
network with average degree 4, and then a second-order phase
transition will be shown with infinite $DL$. For finite $DL$,
however, $\psi_{\infty}$ and $\phi_{\infty}$ behave as the
first-order phase transition characterized by discontinuously
changing from nonzero fraction to zero, which differs from the
second-order phase transition occurred in the first model with
interdependent SF networks. It suggests that enhanced dependence
strength between subnetworks leads to more robust performance and
the change from first-order phase transition to second-order phase
transition. In addition, this simulation result agrees well with
the prediction of Eqs.(\ref{psi}) and (\ref{phi}). In panels (c)
and (d), for power-law dependence relation, similar tendency of
giant components on $p$ is found. Nevertheless, there is a little
deviation between the prediction and the simulation in the case of
$DL=1$. The actual degree distribution of dependence degrees in
simulation has fat-tail deviating from the distribution predicted
by the theory, which causes that nodes with large degree in
fat-tail, or with more dependence links make them still functional
under larger fraction of nodes randomly attacked in subnetwork A.
So this possibly results in the critical value $p_c$ in simulation
is smaller than that in prediction. In addition, as $DL$
approaches to infinity, $\phi_{\infty}$ in panels (b) and (d)
converges to a Heaviside step function, $H(p-p_c)$, which
discontinuously changes from one for $p>p_c$ to zero for $p<p_c$
and $p_c=0.25$. The potential explanation for this phenomenon is
that two subnetworks will connect fully with each other as the
dependence links between them increase to infinity (actually
increase up to the size of subnetwork in the simulation). When
$p<p_c$, the giant component of subnetwork A disappears, so
$\phi_{\infty}$ is close to zero, and when $p>p_c$, subnetwork B
is almost a complete network as most of its nodes have dependence
links from subnetwork A, so $\phi_{\infty}$ is close to one.

\begin{figure}[h!]
\includegraphics[scale=.08]{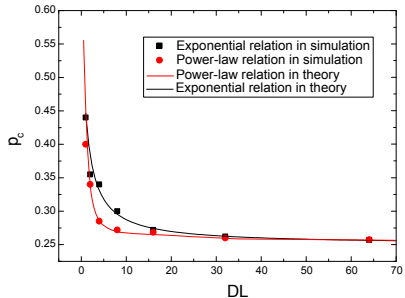}
\centering \caption{(Color online) Comparison of robustness of
networks with two different dependence relations at $N=10^5$ and
$a=b=4$. The critical threshold $p_c$ decreases with the
increasing of $DL$ and approaches the critical value 0.25 of
random percolation of a single network with infinite $DL$. The
value of $p_c$ for power-law relation is always smaller than that
for exponential relation in simulations (symbols), which suggests
that the networks with former relation is more robust to random
failure than networks with latter relation. However, there is an
exception in theoretical result (red line) at $DL=1$ with
power-law relation. Weak dependence strength between subnetworks
may be responsible for this.}\label{ER_pc_DL}
\end{figure}

In order to compare the influence caused by various dependence
relations on the percolation behavior of networks, the critical
threshold $p_c$ as a function of $DL$ is plotted in
Fig.\ref{ER_pc_DL} where the theory is found agreeing well with
the simulation result. As $DL$ increases, the critical threshold
$p_c$ decreases in both relations. In addition, the value of $p_c$
for power-law relation is always smaller than that for exponential
relation with different $DL$ in simulations. In theory, however,
there is an exception in weak dependence strength $DL=1$ where the
critical threshold $p_c$ for power-law relation (marked by red
line) is larger than that for exponential relation (marked by
black line). Similar exception is found in the case of $q\leq0.2$
in Fig.\ref{Pre_Pre_psi_p} in the first model, so it comes to a
strong conclusion that when the dependence strength between
networks is larger than one certain value, the system with
power-law relation behaves more robust against random failure than
networks with exponential relation.

\section{Conclusions and discussion}
In present work, based on two network models, we developed a
framework for studying the effect of dependence relation and
strength in the percolation of two fully interdependent SF and ER
networks, subject to random attack. The addition of dependence
links with random and preferential attachment between subnetworks
in two stochastic models results in the exponential and power-law
distribution of dependence degree, respectively. For both
dependence relations, we find that, 1) in two interdependent SF
networks, strong dependence strength makes the system more
vulnerable and the system only goes through the second-order phase
transition; 2) in two interdependent ER networks, the opposite
results are found that increased dependence strength can enhance
the robustness of system and the system shows a first-order phase
transition. In addition, when the dependence strength is given in
excess of certain value, power-law relation yields greater
robustness than exponential one, which strengthens the known
conclusion that correlated coupled system always has more
robustness than randomly coupled system. The accurate theoretical
analysis needed to be provided to improve this result in future.

The models studied here can help to further understand the design
of real-world interdependent networks where comprise more complex
dependence relations. Through adjusting the parameters of models,
they could also have the flexibility to represent a variety of
interdependent complex systems. Moreover, inspired by
Ref.\cite{hkch}, this work could be extended by taking the
inter-connectivity links between subnetworks into consideration
and more plentiful behavior might be found in the percolation
phase transition of coupled networks.

\end{document}